\documentclass[12pt,oneside]{article}
\usepackage{amsfonts,amssymb,graphicx}
\def\be{\begin{equation}}
\def\ee{\end{equation}}
\def\bea{\begin{eqnarray}}
\def\eea{\end{eqnarray}}
\def\<{\langle}
\def\>{\rangle}
\def\~{\tilde}
\def\s{\sigma}

\def\a{\alpha}
\def\b{\beta}

\def\o{\omega}
\def\t{\tau}

\def\ds{\displaystyle}
\newcommand{\R}{\Bbb R}

\newcommand{\T}{\Bbb T}
\newcommand{\Z}{\Bbb Z}

\newcommand{\av}[1]{\mbox{{\rm Av}}\left(#1\right)}

\newtheorem{theorem}{Theorem}
\newtheorem{lemma}{Lemma}
\newtheorem{definition}{Definition}

\begin{document}
\begin{center}
{\sc\Large
monotonicity and thermodynamic limit for \\short range disordered
models\footnote{Partially supported by Universit\`a di Bologna, Funds for
Selected Research Topics}}
\vskip 1cm
{\bf Pierluigi Contucci\footnote{contucci@dm.unibo.it}, Sandro
Graffi\footnote{graffi@dm.unibo.it}}\\
\vskip 0.5truecm
{\small Dipartimento di Matematica} \\
    {\small Universit\`a di Bologna,
    40127 Bologna, Italy}
    \end{center}
\hfill{\it To Giovanni Jona-Lasinio on his 70-th birthday$\qquad\;\,\,$}
\vskip 1truecm
\begin{abstract}\noindent
If the variance of a short range Gaussian random potential grows like the
volume 
its quenched thermodynamic limit is reached monotonically.
\end{abstract}
\newpage\noindent
{\sc Introduction}\\ \\
The question of the existence of thermodynamic limit  for all standard
(i.e, short
range, see Remark 2 below) models of spin glasses with two-body
in\-terac\-tions has been settled long ago by Khanin
and Sinai \cite{KS} and later generalized to more general interactions by
Zegarlinski \cite{Ze} (previous references on the subject include
\cite{Le},
\cite{Vu}, \cite{PF}). The sharper property of the monotonicity of the free
energy in the
volume  has been proved by van Enter and van Hemmen\cite{VV}. In the
long-range
 case (the most 
important examples being the Sherrington-Kirckpatrick model \cite{MPV}, the
REM
and the GREM \cite{DG}) the existence and the monotonicity has been instead
proved only 
very recently
\cite{GT},\cite{GT2},
\cite{CDGG}. The proof relies on an interpolation argument introduced in
\cite{GT} which has the advantage of
yielding  the subadditivity of the free energy (equivalently,
superadditivity
of the pressure). Exactly as in the ferromagnetic case \cite{Ru}, and in
\cite{VV}, the
subadditivity entails the important property of the {\it monotonicity} of
the
free energy (pressure) as the volume increases.

In this paper we show that the above interpolation argument can be applied
(actually in a slightly different form) to the short range case. For
Gaussian couplings,  in the summable case  and in the non-summable
one as well,   we ge\-neralize  the
Khanin-Sinai and van Enter-van Hemmen results. Namely, for general, mixed,
short-range
$n-$body
interactions ($n$ arbitrary: for the physical relevance of $n>2$ see
e.g.\cite{Li}),   
and assuming free boundary conditions,  the free energy,
internal
energy,
ground state energy are not only bounded but also decreasing
in
the volume. 
Hence their thermodynamic limit is reached monotonically. We remark that in
the
disordered case the monotonicity is even more relevant than in the
ferromagnetic
one because the ground state energy is tacitly assumed monotonic in
all numerical simulations; for a discussion of this point, see
e.g.\cite{Ri}, \cite{BCDG}.

The conclusion which may be drawn by this paper, together with
re\-feren\-ces \cite{GT} and \cite{CDGG}, is that as far as the
thermodynamic
limit is concerned
Gaussian  spin glasses in free boundary conditions
do  not differ from
ordinary ferromagnets: in both cases pressure, internal energy and ground
state
energy are bounded and monotonic in the volume.
\\\\
{\sc Definitions and Examples}\\
Let $M$ be a countable set and consider  a finite subset
$\Lambda\subset M$
of cardinality  $|\Lambda|=N$. To each element $i\in \Lambda$ we
associate a dynamical
variable $\s_i\in{\cal S}\subset \R^k$ (for some fixed integer $k$) equipped
with an {\it a priori} proba\-bility
measure $\nu_i$.  For each $X \subset \Lambda$ we consider
$\s_X=\{\s_i\}_{i\in X}$ and a function
$\Phi_X: \s_X\to \Phi_X(\s_X)\in\R$.\\
In analogy to \cite{Ru} (Sect.2.4, formula 4.3) and \cite{Ze} we  define the
{\it random
potential} as
\be
U_\Lambda(J,\s) = \sum_{X\subset\Lambda}J_X\Phi_X(\sigma_X) \; ,
\label{sgp}
\ee
(with $\Phi_{\emptyset}=0$) under the following assumption:
{\it the coefficients  $J_X$
are independent Gaussian variables with zero mean and variance depending
only on $X$ (and not on $\Lambda)$}
\bea
{\rm Av}(J_X) = 0 \; ,\quad
{\rm Av}(J^2_X) = \Delta^2_X  \; .
\eea
\newpage
{\bf Examples}: \\
\par\noindent
Here  $M=\Z^d$, and $\Lambda$ is a cube.
\begin{enumerate}
\item The Edwards-Anderson model.  ${\cal
S}=\{+1,-1\}$,
$\ds \nu(\s_i)=\frac12[\delta_1+\delta_{-1}]$.
The nearest neighbor case is defined by
$\Phi_{n,n'}(\s_n,\s_{n'})=\s_n\s_n'$
for $|n-n'|=1$,
$\Phi_X=0$ otherwise, and $\Delta^2_X=c^2$. More generally one may consider
a short range 
interaction with with $\Delta^2_X=|n-n'|^{-2d\a}$, $\a>1/2$, or a many-body
interaction with a suitable decay law.
\item Multicomponent spin models (Potts models):  ${\cal S}=\{1,2,...,q\}$,
$\ds \nu(\sigma_i)=
\frac{1}{q}\sum_{l=1}^q\delta_l\;$,
$\Phi_{X}(\s_X)=\delta_{\s_X}$ where $\delta_{\s_X}=1$ if all components of
$\s_X$ 
are equal and zero otherwise.
\item Continuous spin  models: ${\cal S}=\R^k$, $\nu(\sigma_i)=d\mu(x)\geq
0$, 
$\ds \int_{R^k}d\mu(x)=1$
(unbounded case) or
${\cal S}=\T^k$, $\nu(\sigma_i)=d\phi$ (bounded case);
\item Lattice gases: here ${\cal S}=\{0,1\}$,
$\ds \nu(\sigma_i)=\frac12[\delta_0+\delta_{1}]$.
\end{enumerate}
{\bf Remarks}: 
\begin{enumerate}
\item  Of course the examples may be
considered on every
finite dimensional lattice like $\Z^d$ or the triangular lattice etc.
\item The property that $\Delta_X^2$ is independent of the volume
$\Lambda$
characterizes the short range case, such as the
Edwards-Anderson one. In mean field (long range) models, such
as the Sherrington-Kirckpatrick one, the variance has to decrease with $N$
in order to have finite energy density.
\end{enumerate}
Denoting  $P_\Lambda(d\s) \,=\, \prod_{i\in \Lambda}d\nu_i(\s_i)$ we
define:
\begin{enumerate}
\item The random partition function
\be
Z_\Lambda(J) \, := \, \int_{{\cal S}^{N}}P_\Lambda(d\s) e^{U_\Lambda(J,\s)}
\; ,
\ee
\item The random Gibbs-Boltzmann state
\be
\omega(-) :=
\frac{\int_{{\cal S}^{N}}P_\Lambda(d\s) - e^{U_\Lambda(J,\s)}}
{Z_\Lambda(J)} \; ,
\label{omega}
\ee
\item The quenched state
\be
<-> \, := {\rm Av} (\omega(-)) \; ,
\ee
\item The quenched pressure
\be 
P_\Lambda := \, \av{\ln Z_\Lambda(J)} \; .
\label{fe}
\ee
\item
The quenched potential
\be 
U_\Lambda :=  <U_\Lambda(J,\sigma)>
\label{ee}
\ee
\end{enumerate}
We remind that the free energy  $F_\Lambda $ is $-\beta^{-1}P_\Lambda $, and
the internal
energy $E_\Lambda$ is $\beta^{-1}U_\Lambda$.
\\
\\  
 {\sc Superadditivity}
\begin{lemma} 
\be
<J_X\Phi_X> \,\,\, \ge \, 0 \; .
\label{sega}
\ee
\end{lemma}
Proof.\\
We remind the integration by parts for Gaussian variables
\be\label{ibp}
\av{J_Xf(J)} \, = \, \Delta_X^2\av{\frac{df(J)}{dJ_X}} \; ,
\ee
and the correlation derivative formula
\be\label{cd}
\frac{d\omega(\Phi_X)}{dJ_X} \, = \, \omega(\Phi^2_X) - \omega(\Phi_X)^2 \;
\ge 0.
\ee
By applying successively (\ref{ibp}) and (\ref{cd}) we obtain
\bea\nonumber
<J_X\Phi_X> \; &=& \, \av{J_X\omega(\Phi_X)}\; = \\
&=& \Delta^2_X \av{\omega(\Phi^2_X) - \omega(\Phi_X)^2} \ge 0
\; .
\label{se}
\eea
As a corollary of lemma 1 we have
\be
<U_\Lambda(J,\s)> \, = \, \sum_{X\subset \Lambda} \Delta^2_X
\av{\omega(\Phi^2_X) - \omega(\Phi_X)^2} \ge 0
\label{inte}
\ee
\begin{definition} 
Consider a {\it partition} of $\Lambda$ into $n$ non empty disjoint sets
$\Lambda_s$:
\be
\Lambda=\bigcup_{s=1}^{n}\Lambda_s \; ,
\ee
\be
\Lambda_s\cap\Lambda_{s'} = \emptyset \; .
\ee
For each partition the potential generated by all interactions among
different subsets is defined as
\be
{\tilde U_\Lambda} = U_\Lambda - \sum_{s=1}^{n}U_{\Lambda_s} \; ;
\ee
from (\ref{sgp}) we have that
\be
{\tilde U_\Lambda} = \sum_{X\in {\cal C}_\Lambda} J_X\Phi_X
\ee
where ${\cal C}_\Lambda$ is the set of all $X\subset\Lambda$ which are not
subsets of any $\Lambda_s$.
\end{definition}
\begin{theorem}
The quenched potential is superadditive:
\be
<{U_\Lambda}> \; \ge \, \sum_{s=1}^{n}<U_{\Lambda_s}>
\ee
\end{theorem}
Proof. Direct consequence of (\ref{sega}). In fact:
\be
<{\tilde U_\Lambda}> \, = \, \sum_{X\in {\cal C}_\Lambda} <J_X\Phi_X>  \, =
\, 
\sum_{X\in {\cal C}_\Lambda}
\Delta^2_X \av{\omega(\Phi^2_X) - \omega(\Phi_X)^2} \ge 0 \, .
\ee
\begin{theorem}
The quenched pressure is superadditive:
\be
P_\Lambda \, \ge \,  \sum_{s=1}^{n}P_{\Lambda_s} \; .
\label{super}
\ee
\end{theorem}
Proof.\\
To each partition of $\Lambda$ we associate the interpolating potential for
$0\le t\le 1$
\be
U_\Lambda(t) \, = \, \sum_{s=0}^{n}
\sqrt{t_s}U_{\Lambda_s}^{(s)} \, ,
\ee
with $t_0=t$, $t_s=(1-t)$ for $1\le s\le n$, $U^{(0)}_{\Lambda_0} =
U_\Lambda$ and
\be
U_{\Lambda_s}^{(s)} \, = \, \sum_{X\subset\Lambda_s} J^{(s)}_X\Phi_X\; ,
\label{local}
\ee
where any $J^{(s)}_X$ is a centered independent Gaussian
\be
\av{J^{(s)}_XJ^{(q)}_Y}=\delta_{s,q}\delta_{X,Y}\Delta^2_X
\ee
(the symbol Av is here the average with respect to all the J's).
We define the interpolating partition function
\be
Z_\Lambda(t) \, = \, \int_{{\cal S}^{N}}P_\Lambda(d\s) e^{U_\Lambda(t)} \; ,
\ee
and we observe that
\be
\label{ifa}
Z_\Lambda(0) \, = \, \prod_{s=1}^{n}Z_{\Lambda_s}(J^{(s)}) \; , \; \quad
Z_\Lambda(1)=Z_{\Lambda}(J)\; .
\label{st}
\ee
Consider the interpolating pressure
\be 
P_\Lambda(t) := \, \av{\ln Z_\Lambda(t)} \; ,
\label{fe}
\ee
and the corresponding states $\o_t(-)$ and $<->_t$.
Thanks to  (\ref{st}) we get
\be
\label{ifa}
P_\Lambda(0) \, = \, \sum_{s=1}^{n}P_{\Lambda_s} \; , \; \quad
P_\Lambda(1)=P_{\Lambda} \; .
\ee
We observe now that
\be\label{derif} 
\frac{d}{dt} P_\Lambda(t) \,=\,\sum_{s=0}^{n}
\frac{\epsilon_s}{\sqrt{t_s}}<U_{\Lambda_s}^{(s)}>_t
\; , 
\ee 
with $\epsilon_0=1$ and $\epsilon_s=-1$ for $1\le s\le n$. For each $s$ we
have
\be
<U_{\Lambda_s}^{(s)}>_t \, = \, \sum_{X\subset\Lambda_s} <J^{(s)}_X\Phi_X>_t
\; ;
\label{loc}
\ee
Integrating by parts  each addend we obtain
\be
<J^{(s)}_X\Phi_X>_t \, = \, \sqrt{t_s}\Delta_X^2 \av{\omega_t(\Phi^2_X) -
\omega_t(\Phi_X)^2}\,
\ee
and summing up all the contributions in (\ref{derif}):
\bea
\nonumber
\label{fino}
\frac{d}{dt} P_\Lambda(t) \, &=& \, \sum_{X\subset\Lambda}\Delta_X^2
\av{\omega_t(\Phi^2_X) - \omega_t(\Phi_X)^2}
- 
\\
\nonumber
&-& \sum_{s=1}^{n}\sum_{X\subset\Lambda_s}\Delta_X^2
\av{\omega_t(\Phi^2_X) -
\omega_t(\Phi_X)^2}\; \\
&=&\sum_{X\in {\cal C}_\Lambda}
\Delta_X^2 \av{\omega_t(\Phi^2_X) - \omega_t(\Phi_X)^2}\ge 0 \; .
\eea
>From (\ref{ifa}) and (\ref{fino}) we immediately get formula
(\ref{super}).\\\\
{\sc Boundedness}\\
For any random potential we define the quantity
\be
||U|| \, = \, \sup_{\Lambda}\frac{1}{N}\av{U_\Lambda(J,\s)^2} \, = \,
\sup_{\Lambda}\frac{1}{N}\sum_{X\subset\Lambda}\Delta^2_X\Phi^2_X \, .
\label{bnd}
\ee
Potentials with a finite $||U||$ are called stable.
\begin{theorem}
A stable random potential admits an internal energy and a
quenched pressure bounded by the volume.
\end{theorem}
{\bf Proof.} By (\ref{inte}):
\be
<U_\Lambda(J,\s)> \, = \, \sum_{X\subset \Lambda} \Delta^2_X
\av{\omega(\Phi^2_X) - \omega(\Phi_X)^2} \le 2||U||N \, .
\ee
Using the Jensen inequality
\bea\nonumber
P_\Lambda \, = \, \av{\ln Z_\Lambda(J)} \, &\le& \, \ln\av{Z_\Lambda(J)}\,
=\; \\\nonumber
=\ln\int_{{\cal S}^{N}}P_\Lambda(d\s)
\av{e^{U_\Lambda(J,\s)}}&=&\ln\int_{{\cal S}^{N}}P_\Lambda(d\s)
e^{\frac{1}{2}\sum_{X\subset\Lambda}\Delta^2_X\Phi^2_X}=\\
&\le& \frac{1}{2}||U||N
\eea
As a consequence for finite $||U||$ one has
\be
\sup_{\Lambda} \frac{1}{N} U_\Lambda \le \infty \; ,
\ee
and
\be
\sup_{\Lambda} \frac{1}{N} P_\Lambda \le \infty \; .
\ee
\vskip 0.3cm\noindent
{\sc Thermodynamic Limit}\\
Let us verify the stability condition in the above examples.
\begin{enumerate}
\item Edwards-Anderson.\\
For the nearest neighbor case
\be
\sum_{(n,n')}\Delta^2_X = 2dNC^2
\ee
\item
More generally for the short range case with $\a>1/2$
\be\label{ns}
\sum_{n,n'}\Delta^2_X = \sum_{n,n'} \frac{1}{|n-n'|^{2d\alpha}} \le {\rm
const\,} N 
\ee
\end{enumerate}
By Theorems 1 and 2 the previous models have an internal energy
per particle and a free energy per particle which exist in the
thermodynamic limit.\\\\
{\bf Remarks}:
\begin{enumerate}
\item We point out that for short range models we only need to impose
the boundedness condition (\ref{bnd}) while the superadditivity always holds
thanks to the condition of independence of the variance $\Delta_X^2$ from
the volume (see also
the remark 2). In the mean field case
the variance of the interactions depends on the volume and subadditivity
is based on an  inequality among the covariances \cite{CDGG}
\be
N_1c_{N_1}(\s,\t)+N_2c_{N_2}(\s,\t)-Nc_N(\s,\t)\ge 0 \, .
\ee
One may check that such an inequality reduces, in the short range case, to
the positivity of the
right hand side of (\ref{fino}).
\item Our result may be extended in two directions by exactly
the same procedure of \cite{GT}. First one can prove by standard probability
arguments
that the above
statement entails the almost sure convergence of pressure and ground state
energy per particle.
Second our result may be extended to non Gaussian $J$ (see section 4.2 of
\cite{GT} and \cite{T}):
if  $J_X$ is for all $X$ an even random variable with a finite 4th moment
the
integration by parts
(\ref{ibp}) is replaced by the more general formula
\be
{\rm 
Av}(J_XF(J))={\rm 
Av}(J_X^2F'(J))-\frac{1}{4}{\rm Av}(|J_X|\int_{-|J_X|}^{|J_X|}(J^2_X-x^
2)F'''(x)dx) \, .
\ee
When used within  Theorem 2 it generates in formula (\ref{fino}) a
correction of
order $O(\sqrt{N})$. Once the density is taken the correction vanishes in
the
thermodynamic limit. In
general,  however, we cannot establish its sign and the monotonicity is
lost.
\item 
The present result holds for free boundary conditions.
In general it is proved\cite{Ze}
by standard surface over volume arguments that the quenched quantities are
independent
of the boundary conditions, but  the monotonicity property is lost.
\item It is interesting to observe that the interpolating strategy
does apply also to standard ferromagnetic systems. Consider for instance the
d-dimensional ferromagnetic Ising model with nearest neighbor Hamiltonian
$H_\Lambda(\s)=-\sum_{(n,n')}\s_n\s_{n'}$. An interpolating functional would
be
\be
\a(t) \, = \, \log 
\sum_{\s}e^{-\b[tH_\Lambda(\s)+(1-t)\sum_{s=1}^{n}H_{\Lambda_s}(\s)]} \, .
\label{obv}
\ee
An easy calculation which goes parallel to Theorem 2 yields
\be
\frac{d\a}{dt}(t) \, = \, \sum_{(n,n')\in {\cal C}} \o_t(\s_n\s_{n'}) \, .
\label{ferro}
\ee
Since $0 \le t\le 1$ the t-interaction on (\ref{obv}) is still ferromagnetic
and 
the Griffiths inequality (see for instance \cite{Ru})
$\o_t(\s_n\s_{n'})>0$ gives the positive sign of the former expression
ensuring 
the monotonicity of the limit.
\end{enumerate}
{\bf Acknowledgments}. We thank A.Bovier, A.C.D.van Enter, M.Degli
E\-sposti,
C.Giardin\`a, F.den Hollander, F.Guerra, I.Nishimori, E.Olivieri,
F.L.To\-ninelli for interesting discussions. P.C.
thanks the Tokyo Institute of
Technology for the kind hospitality.

\end{document}